\documentclass[twocolumn,prd,showpacs,preprintnumbers,amsmath,amssymb,aps,float,floatfix]{revtex4}
\usepackage{graphicx}
\usepackage{amsmath}
\usepackage{amsfonts}
\usepackage{amssymb,amsthm}
\usepackage{mathrsfs}
\usepackage{epsfig}
\usepackage{color}
\begin{document}
\title{Boson stars: Chemical potential and quark condensates}
\author{Jitesh R. Bhatt}
\email{jeet@prl.res.in}
\author{V. Sreekanth}
\email{skv@prl.res.in}
\affiliation{Physical Research Laboratory, Ahmedabad
380009, India}
\begin{abstract}
We study equilibrium solutions of a boson star in mean field
approximation using a general relativistic framework. 
Dynamics of the constituent matter in such a star is
described by a scalar field.
The statistical aspect of the matter equilibrium is investigated by incorporating 
the effect of finite chemical potential in equation of state. 
We analyze the two possible situations where  the scalar field is either a
composite one -suitable for describing diquark-condensates or 
a fundamental. We derive a generalized set of Tolman-Oppenheimer-Volkov (TOV)
equations to incorporate the metric dependence of chemical potential in 
general relativity. It is demonstrated that the introduction of finite chemical
potential can lead to a new class of the solutions where the maximum mass and 
radius of a star change in a significant way. In general the density profile
has node like structures due to introduction of finite chemical potentials.
It is shown that these nodes can be avoided by applying
constraints on the values of the central pressure and central chemical 
potential. This in turn can reduce the parameter space available for the
stable solutions.
We also discuss the case when the boson star is made of diquark condensates.
It is shown that when the self-interaction of the condensate is negligible,
the typical energy density of the star is too high to have a boson star of diquark
condensates. Our results indicate, even after the effect of self-interaction is
incorporated, the stable equilibrium can occur only in a low-density and
high-pressure regime.
\end{abstract}

\maketitle
\section{INTRODUCTION}
The boson stars are very fascinating objects as their self-gravity is not
balanced by the degeneracy pressure like the other compact stars such as  
a white dwarf or a neutron star, but the Heisenberg uncertainty play a crucial
role in their stability (see Ref. \cite{sch} for a recent 
review). Wheeler was first to consider the boson star in 1955 when he studied the
stable equilibrium generated by the self-gravitating photons called {\it geons}
\cite{wheel}. Later Kaup analyzed a new class of boson stars as a self-consistent
solutions of Einstein-Klein-Gordon equations in Ref. \cite{kaup}. Since then the most
of the boson star studies focus on the the stars made of self-gravitating scalar-field
\cite{sch, rev2}. This is because there are good reasons to believe that there exists
a some fundamental scalar field in nature even though no experiment has detected
it so far. The solutions found by Kaup in Ref. \cite{kaup} were shown to describe the
ground state of boson stars without using any perfect-fluid approximation or some
approximation to equation of state by Ruffini and Bonazzola \cite{rb}. However in this
work the self-interaction of the scalar-field was ignored. 
Subsequently it was shown by Colpi et. al. \cite{shapiro} that inclusion of even
a very small interaction term can significantly alter the solutions obtained in Refs. 
\cite{kaup,rb}. It turns out that the inclusion of a small interaction term in the
scalar field Lagrangian give rise to an additional pressure that would help the
Heisenberg uncertainty to counter the self-gravity. This results in 
a boson star with much larger  mass than ones reported in Refs. \cite{kaup,rb}. 
Also the issues of dynamical stability  of boson stars have been rigorously analyzed by
applying perturbative methods \cite{stab}. Also are studies that show that there
can be a boson star in the galactic center \cite{gallaxy-bs}. 
At present a very intriguing situation persists in the field: on the one hand there 
is no observational evidence which can indicate that a boson star exists, on the other hand 
their existence seem to be a
reasonable consequence of well-tested physical theories like general relativity
and Klein-Gordon equations.
This should be a good enough reason, in our opinion, to further investigate the
properties of boson stars.

Boson stars with a repulsive self-interaction mediated by vector mesons
within the effective theory framework were studied only recently \cite{agni}.
It is found that mass-radius curves satisfy the scaling properties for
arbitrary boson masses and interaction strengths.
There exist an another novel possibility of having a boson star with a composite 
scalar field. Such a composite scalar field can possibly arise when the superconducting
colour quark matter undergo a transition from the BCS phase to Bose-Einstein
condensation phase via a so called BCS-BEC transition \cite{pir1, bec}.
It should be noted that BCS regime of the colour superconducting phase is well studied
using the techniques of perturbative QCD. But in BCS-BEC crossover 
can occur in low density and low temperature regime, where the coupling constant is
large and the perturbative techniques are not available. One may use the methods
of effective theory to describe the matter in the strongly coupled regime.
In this picture, instead of a repulsive vector-meson exchange
one may have repulsive self-interaction of the scalar field describing the (diquark)
condensates \cite{pir1}. 
A diquark-BEC phase may occur at low density regime somewhere
between colour neutral nuclear matter and the free quark phases. This requires
an additional constraint namely - the size of the condensates or the coherent
length $\xi$ should not be larger than the inter-particle spacing i.e. $n^{-1/3}$
, where $n$ is the number density.  

In this work we analyze the equilibrium solutions of a boson star in a low temperature
finite chemical potential regime. As far as we know boson stars  with a finite chemical potential 
are not properly considered in the literature. Majority of study do not focus on the
statistical aspect of the self-gravitating boson gas but they rather appeal to
find a specific equilibrium (quantum) solution of the Einstein-Klein-Gordon system. 
In the context of quark star we would like to mention that possibility of
diquark star has already been considered in the literature \cite{diquarkstar}.
However the effect of Bose-Einstein condensation was not considered in
this work. This effect can be significant in a low temperature finite
chemical potential regime \cite{pir1}. Moreover, in general relativity 
temperature and chemical potential are local functions of space and time and
therefore they depend upon the metric. The metric dependence of these
quantities can be specified by using the well-known Tolman conditions for a thermal 
equilibrium in the gravitational field \cite{tol34, bilic99}. Again this aspect was not 
considered in the literature on boson or diquark stars. In this work we derive a generalized
set of Tolman-Oppenheimer-Volkov equations by incorporating the effect of metric dependence
of chemical potential. This has resulted in an additional differential equation describing
the space-time evolution of the chemical potential coupled with the TOV \cite{tov} equations.

In a very high density system such as a boson star, the introduction of finite chemical
potential can rise a question. However in the standard model, for example, quark numbers are 
conserved and one can introduce chemical potential for them. In addition for a 
BCS-BEC crossover kind of transitions chemical potential play a very important role
\cite{pir1}. Chemical potentials are known to alter the global minimum of a scalar-field. 
In this work have demonstrated that when the scalar field mass $m$ is comparable
to the chemical potential $\mu$, the mass-radius relations and the maximum mass of
the star are significantly changed. 

In what follows we first derive the generalized set of TOV equations and study their numerical solutions 
by providing the equation of state as an input. 
The equation of state is derived in the mean-field approximation. We
use for the parameter values of the diquark phase in a manner consistent with that
given in Ref. \cite{pir1}.

\section{FORMALISM}

Lagrangian density for the scalar-field can be written as
\begin{equation*}
 \mathcal{L} = - \sqrt{|g|}~ \mathcal{L}_\Phi
\end{equation*}
with
\begin{eqnarray}\nonumber
 \mathcal{L}_\Phi = &-&\left[ g^{\mu\nu}\partial_\mu \Phi^\dagger\partial_\nu \Phi + i\mu g^{00}\delta_0^\alpha 
\left(\Phi^\dagger \partial_0 \Phi - \Phi \partial_0 \Phi^\dagger\right)\right] \\ 
                  &-& U(|\Phi|^2)\label{lagrangian}
\end{eqnarray}
\noindent
where, the potential $U(|\Phi|^2)$ is defined as
\begin{equation}\label{potential}
U(|\Phi|^2) = (m^2 + g^{00} \mu^2) |\Phi|^2 + ~g |\Phi|^4.
\end{equation}
\noindent
The energy-momentum tensor $T_{\mu\nu}(\Phi)$ written from this as follows:
\begin{eqnarray}
T_{\mu\nu}&=&\left[\partial_\mu \Phi^\dagger \partial_\nu \Phi+\partial_\mu \Phi\partial_\nu \Phi^\dagger\right] \\ \nonumber
&+& i\mu g^{00}g_{0\mu} 
\left[\Phi^\dagger \partial_\nu \Phi - \Phi \partial_\nu \Phi^\dagger\right]+g_{\mu\nu}\mathcal{L}_\Phi
\end{eqnarray}

The equations of motion of the scalar field is given by
\begin{equation}
 \Box + 2i\mu g^{00}\partial_0+\frac{dU}{d|\Phi|^2}=0
\end{equation}
where 
\begin{equation}
 \Box := -\frac{1}{ \sqrt{|g|}} \partial_{\mu}(\sqrt{|g|}g^{\alpha \mu}\partial_{\alpha})
\end{equation}

We are considering the 'standard' form of a spherically symmetric and static space-time metric \cite{wein} 
given by
\begin{equation}\label{metric}
 ds^2 = -B(r) dt^2 + A(r) dr^2 + r^2 (d\theta^2 + \sin^2\theta d\phi^2),
\end{equation}
where, we have used the units in which speed of light $c=1$. Next to study the equilibrium configuration, we use the stationarity {\it ansatz} 
$\Phi(r,t)=\phi(r) e^{-i\omega t}$. Now the energy density $\rho$ and pressure $p$ can be obtained from the energy-momentum tensor:
\begin{equation}\label{rho}
 \rho = \frac{T_{tt}}{B} = \frac{(\omega^2 - \mu^2)}{B}~|\Phi|^2 +~ m^2~|\Phi|^2+g~|\Phi|^4
\end{equation}
and
\begin{equation}\label{p}
 p = \frac{T_{rr}}{A} = \frac{( \omega + \mu )^2}{B} ~|\Phi|^2 - ( m^2~|\Phi|^2+g~|\Phi|^4).
\end{equation}

And finally the equation of motion for the scalar field gives,
\begin{equation}\label{disp}
 \frac{( \omega + \mu )^2}{B} = m^2 + 2g~|\Phi|^2 . 
\end{equation}

\noindent
It must be noted here that while writing equations (\ref{rho}-\ref{disp}) we have 
dropped the spatial derivative terms of the scalar field. It can be shown that $\frac{d\phi}{dr}$ 
is of the order of $\Lambda^{-1}$. Here $\Lambda=\frac{gM^2_{planck}}{2\pi m^2}$ 
is an extremely large number ($>10^{30}$) for the parameter regime of the interest.
This is equivalent to the mean field approximation ($\partial_r \Phi=0$). 
Under this condition one can use the TOV equation to describe a self-gravitating bose star 
\cite{shapiro}.

Using equations (\ref{rho}-\ref{disp}) one can rewrite $p$ and $\rho$ 
\begin{eqnarray}\label{ptilde}
 p = g~|\Phi|^4\\ \label{eos}
\rho = \rho_0 \left[ 4\left( 1-\tilde{\mu}\sqrt{1+\sqrt{\tilde{p}}}\right)\sqrt{\tilde{p}} + 3 \tilde{p} \right]  
\end{eqnarray}
\noindent
where, $\rho_0 = \frac{m^4}{4g}$, $\tilde{\mu} = \frac{\mu / m}{\sqrt{B}}$ and $\tilde{p} = p / \rho_0$
is the dimensionless pressure.
Equation (\ref{eos}) is equivalent to the EOS considered in Refs. \cite{sch, shapiro, ryan} in 
the limit $\mu=0$.  In the limit of vanishing self-interaction i.e $g\rightarrow 0$,
we have ${p}=0$. In this limit self-gravity of the star will be balanced by the
effect of Heisenberg uncertainty principle only \cite{kaup, lieb}. 
For high values of the $\tilde{p}$ the last term in parenthesis of equation (11) dominates
over the other terms, the EOS approach the relativistic limit i.e. $\tilde{\rho}=3 \tilde{p}$.

The form
of functions $A(r)$ and $B(r)$ in equation (\ref{metric}) can be determined from the  Einstein's field equations,
\begin{equation}\label{GR}
 \mathit{R}_{\mu \nu} - \frac{1}{2} g_{\mu \nu}\mathit{R} = -{8\pi G}T_{\mu \nu}.
\end{equation}

With the metric defined previously we get the relevant equations from (\ref{GR}) as,
\begin{eqnarray}
 \frac{A'}{r A^2}+(1-\frac{1}{A})~\frac{1}{r^2}~&=&~8\pi G\left(\frac{T_{tt}}{B}\right)\\
\frac{B'}{r AB}-(1-\frac{1}{A})~\frac{1}{r^2}~&=&~8\pi G\left(\frac{T_{rr}}{A}\right),
\end{eqnarray}
where $'$ denotes differentiation with respect to $r$. With $A(r)=(1-2GM(r)/r)^{-1}$, we get the stellar structure equations as,

\begin{equation}\label{B'}
\frac{B'}{B}=\frac{2G}{r^2}\frac{\left[M+4\pi r^3 p\right ]}{(1-\frac{2 GM}{r})},
\end{equation}
\begin{equation}\label{tov2}
\frac{dM}{dr}= 4\pi r^2 \rho(r).
\end{equation}
Hydrostatic equilibrium expressed as \cite{wein},
\begin{equation}\label{hydro}
 \frac{B'}{B}=-\frac{2p'}{p+\rho}
\end{equation}
can be used to rewrite the equation (\ref{B'}) as 
\begin{equation}\label{tov1}
 p'=-\frac{G}{r^2}\frac{\left[\rho+p\right ]
\left[M+4\pi r^3 p\right ]}{(1-\frac{2 GM}{r})}
\end{equation}

The equations (\ref{tov2}) and (\ref{tov1}) are known as the TOV equations \cite{tov}, 
which has to be solved together with the equations of state. 
But in our case $\rho$ is a function of chemical potential $\mu$, which in turn depends on the metric. 
Metric dependence of $\mu$ can be understood using equation (\ref{hydro}) and thermodynamic Gibbs-Duhem relation,
\begin{equation}
 d\frac{p}{T}~=~n~d\frac{\mu}{T}-\rho~d\frac{1}{T}
\end{equation}
and can be written as \cite{bilic99},
\begin{equation}
 \frac{B'}{B}=-\frac{2\mu'}{\mu}.
\end{equation}
Finally the we get the desired relation with the help of equation (\ref{B'}) as,
\begin{equation}\label{tov3}
\mu'= -\mu \frac{G}{r^2}\frac{\left[M+4\pi r^3 p\right ]}{(1-\frac{2 GM}{r})}.
\end{equation}
So \textit{generalized stellar structure equations in presence of chemical potential} are given by 
equations (\ref{tov2}),(\ref{tov1}) and (\ref{tov3}). 
These equations are to be solved simultaneously with equation(\ref{eos}), to find out the properties of 
stellar configurations.

\section{RESULTS AND DISCUSSION}

Before solving stellar structure equations numerically let us consider the following
qualitative arguments. 
First consider the case when there is no self-interaction term and also no chemical-potential
i.e. $g=0$ and $\mu=0$. In this case the critical mass and radius of the star given 
$R\sim \frac{1}{m}$ and $M=M_{planck}^2/m$ \cite{kaup}.
Next, if the boson mass is around $100 GeV$, the corresponding
critical mass of the star would be  $M=10^{17} M_{planck}\sim 10^9 kg\sim 10^{-21}M_N$
and its radius $R\sim10^{-22} R_N$ where,$M_N$ and $R_N$
are mass and radius of a neutron star. Consequently, the  density of 
such a boson star would be $10^{45}$ times that of neutron star. Thus for a diquark-condensate
with mass $(0.5-1)GeV$ \cite{pir1}, the density of the corresponding star would be
$10^{41}$ times of that of a neutron star. This is an extremely large value of
the matter density for a diquark star to exist. However, things can change
if the self-interaction term is nonzero. It is known that for the case $g\neq 0$ (and $\mu=0$), the scalar-field
$\mid\Phi\mid$ scales as $M_{Planck}/\Lambda$ if
$\Lambda=\frac{gM^2_{planck}}{2\pi m^2} \gg 1$\cite{shapiro}. This condition is  to satisfy even for a very small values
of $g$. Comparing the quadratic and quartic  terms in the potential $U$, we get 
$\frac{g\mid\Phi\mid^4}{m^2}\sim \Lambda\frac{\mid\Phi\mid^2}{M^2_{planck}}\sim O(1) $.
The energy density $\rho$ will be  about $ m^2M^2_{planck}/\Lambda$.
Thus the energy density would corresponds to that of a non-interacting boson if mass is
rescaled as $ m\rightarrow m  \sqrt{\frac{1}{\Lambda}}$. 
From this the maximum mass can be found \cite{kaup, maxmass} to be
\begin{equation}\label{mcrit}
M_0\approx \frac{2}{\pi}\sqrt{{\Lambda}}
\frac{M^2_{planck}}{m}
\end{equation}
\noindent
This is also matches with the estimate critical mass obtained numerically in Ref. 
{\cite{shapiro}. 
In our case, in the flat space, the classical potential has non-trivial minimum at 
$\mid\Phi\mid^2=\frac{\mu^2-m^2}{2g}$. We have assumed that $m^2>0$ and for $\mu^2>m^2$
we have condensation. In this case the mass scaling will be different and it is
given by $m\rightarrow m/\sqrt{\Lambda}\left(\frac{\mu^2}{B}-1\right)$ and the
maximum mass may be given by $\sqrt{\Lambda}\frac{M^2_{planck}}{m}\left(\frac{\mu^2}{B}-1\right)$.
This suggest that the maximum mass should decrease if the chemical potential is decreasing. Since
the chemical potential term here depend on space-time the above is only a rough estimate
of the maximum mass. It also must be noted that the above arguments are not valid if
$\frac{(\mu/m)^2}{B}<1$.

The validity condition $\xi < n^{-1/3}$ for the condensate
description. $\xi(=\frac{1}{\Delta})$ can be found from the gap equation \cite{raja}
\begin{equation}
\Delta = 2\mu exp\left( -\frac{3\pi^2}{\sqrt{2}g}\right),
\end{equation}
\noindent
for $\mu\approx 400-500 MeV$ and the  corresponding running coupling constant
$3.56$. However, at this juncture  the above equation should not
be taken too seriously as it is based upon the arguments of the perturbation
theory. The BCS-BEC transition is supposed to take place  in the high coupling
regime. The quark-condensation condition in the flat-space is  ${(\mu/m)^2}>1$
\cite{pir1}. However, in the case of a boson star it needs to be locally satisfied.

Next,  we write the generalized TOV equations  in a dimensionless form. We have already introduced the 
dimensionless variables for the pressure $\tilde{p}$, density $\tilde{\rho}$ and chemical potential 
$\tilde{\mu}$. We calculate the mass $M$ in the dimensionless unit $\tilde{M} = M/M_o$ where $M_o$ is 
the solar mass. Radius of the star ($R$) is also computed in terms of $\tilde{R} = R/R_o$, with 
$R_o = G M_o = 1.47 km$. Radius of the star can be defined as the distance from the center 
where $\rho$ becomes zero. We solve equations (\ref{tov2}, \ref{tov1}, \ref{tov3}) by specifying various
values of the central pressure $p_c$ and the chemical potential $\mu_c$, while $M$ at the center is
considered to be zero.

From the form of equation (11) it is clear that, the energy density could be
negative if the values $p_c$ and $\mu_c$ are chosen arbitrarily. Therefore it is required to constrain
their values by imposing the condition $\tilde{\rho} > 0$.\\

\paragraph*{$\tilde{\rho} > 0$ condition and two branches of solutions}: 
\begin{figure}
\includegraphics[width=7cm,height=7cm]{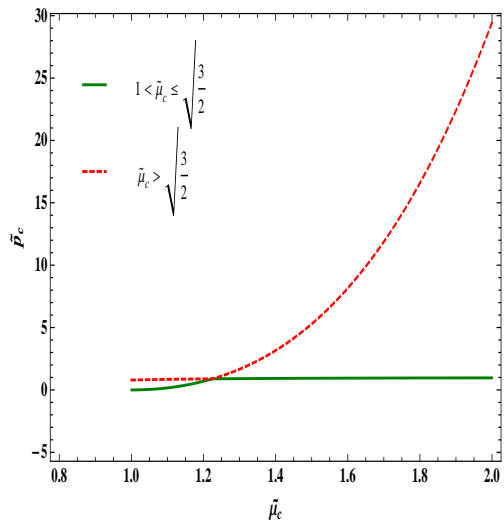}
\caption{The graph shows the solutions for initial conditions of pressure $\tilde{p}_c ~\text{and chemical potential} ~
\tilde{\mu}_c$ for 
$\tilde{\rho}_c\geq 0$.}\label{branches}
\end{figure}

The condition (on the initial conditions) implies that there are following two branches 
of the solutions: 

\begin{eqnarray}
\tilde{p}&\geq& \frac{16}{81} \left(9-15 \tilde{\mu} ^2+8 \tilde{\mu} ^4\right)\\
\nonumber
&-&\frac{32}{81} \sqrt{-27 \tilde{\mu} ^2+72 \tilde{\mu} ^4-60 \tilde{\mu} ^6+16 \tilde{\mu} ^8};~
(1<\tilde{\mu} \leq \sqrt{\frac{3}{2}})\\
\tilde{p}&\geq& \frac{16}{81} \left(9-15 \tilde{\mu} ^2+8 \tilde{\mu} ^4\right)\\
\nonumber
&+&\frac{32}{81} \sqrt{-27 \tilde{\mu} ^2+72 \tilde{\mu} ^4-60 \tilde{\mu} ^6+16 \tilde{\mu} ^8};~(\tilde{\mu} >\sqrt{\frac{3}{2}})
\end{eqnarray}

 The solid curve in figure (1) represents the first branch given by equation (25), which is valid the values of 
chemical  potential defined in the range $1 <\tilde{\mu} \leq \sqrt{\frac{3}{2}}$. The curve with the dotted 
line represents the branch given by equation (26). From figure (1) it is clear that solutions of the
equation (25) is valid even in the region $1 <\tilde{\mu} \leq \sqrt{\frac{3}{2}}$.

\begin{figure}
\includegraphics[width=7cm,height=7cm]{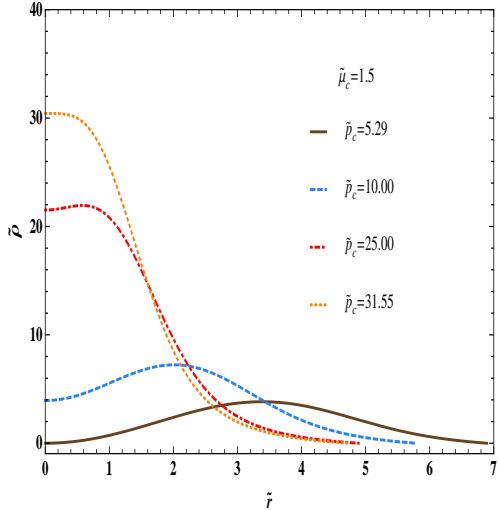}
\caption{For a given chemical potential the density profile can have nodes. The upper most
curve is obtained by imposing the no-node condition.} 
\end{figure}

 Figure (2) shows the density profile as function of $\tilde{r}$ for various values of the central
pressure and the chemical potential $\tilde{\mu}=1.5$. The solution indicates the nodes in the
profiles. However, this is a unphysical behavior and it  can be avoided by imposing
the no-node condition on $\rho$.\\

\paragraph*{No-Node condition}: 
We need to make sure that $\tilde{\rho}'(r=0) < 0$ inside the star. Analytically this condition can be represented as,
\begin{eqnarray}
\left(2+3 \sqrt{\tilde{p}(0)}\right) \left(\sqrt{1+\sqrt{\tilde{p}(0)}}-\tilde{\mu }(0)\right) \tilde{p}'(0)\\
\nonumber
-4 \left(1+\sqrt{\tilde{p}(0)}\right) \tilde{p}(0) \tilde{\mu }'(0)<0
\end{eqnarray}

The upper most curve in figure (2) is plotted by incorporating the no-node condition on the initial
values of pressure and chemical potential. Thus the implementation of the no-node condition shows
that for a given chemical potential there exist a lower bound on the central pressure. If one specify 
the central pressure below this bound then the nodes develops in the density profiles. Our numerical
study indicates that if the initial values of the pressure and chemical potential are chosen just above
the two branches of '$\tilde{\rho}> 0$ condition' as shown in figure (1), 
the density profile develops the nodes. One can avoid the nodes for higher values of the central
pressure $p_c$, but this will also increase the central density as implied by equation (\ref{eos}).
It should be noted that the nodes in our analysis arise solely due to the introduction of the
chemical potential and therefore their origin is different than the nodes in the wave functions
reported in earlier literature of boson-star (for example \cite{shapiro}).

\begin{figure}
\includegraphics[width=7cm,height=7cm]{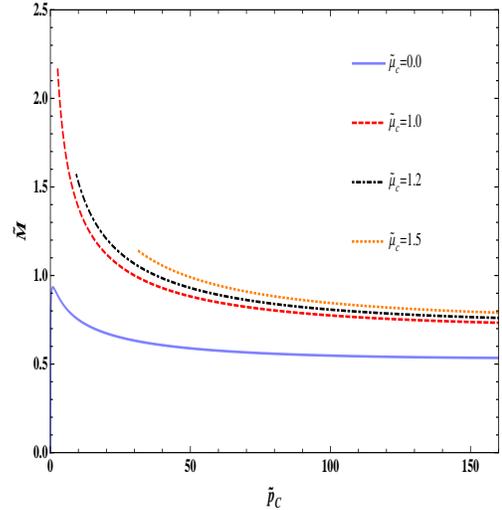}
\caption{Maximum Mass of the star versus initial pressure for different values of $\tilde{\mu}_c$ is plotted. 
The peak of the graph in each case denotes the highest maximum mass configuration.}
\end{figure}

Figure (3) shows graph of  the star-mass as a function  of the central pressure for  different
values of the chemical potential. The lower most curve represents the case of zero chemical-potential.
The mass is increasing with increasing the central pressure (density) and it reaches a plateau.
This behavior is in agreement with the earlier work. Also the value of maximum-mass is in
agreement with Ref. \cite{shapiro}. The introduction of a finite chemical-potential increases the
overall value of the maximum-mass and the plateau. This figure shows that when the values
of  $\mu_c$, increases, the maximum-mass values occurs at higher values of $p_c$. 
The lower values of $p_c$ in this case ($\mu_c>1$) are ruled out due to the no-node condition. 
If this condition is violated then the maximum-mass configurations would occur at much lower $p_c$. 
And in this case the  maximum-mass would be much higher than the $\mu_c=1$ case. 
Thus the consistent solutions with finite  chemical-potential requires to have higher values of the
central-pressure in comparison with the $\mu_c=0$ case.

\begin{figure}
\includegraphics[width=7cm,height=7cm]{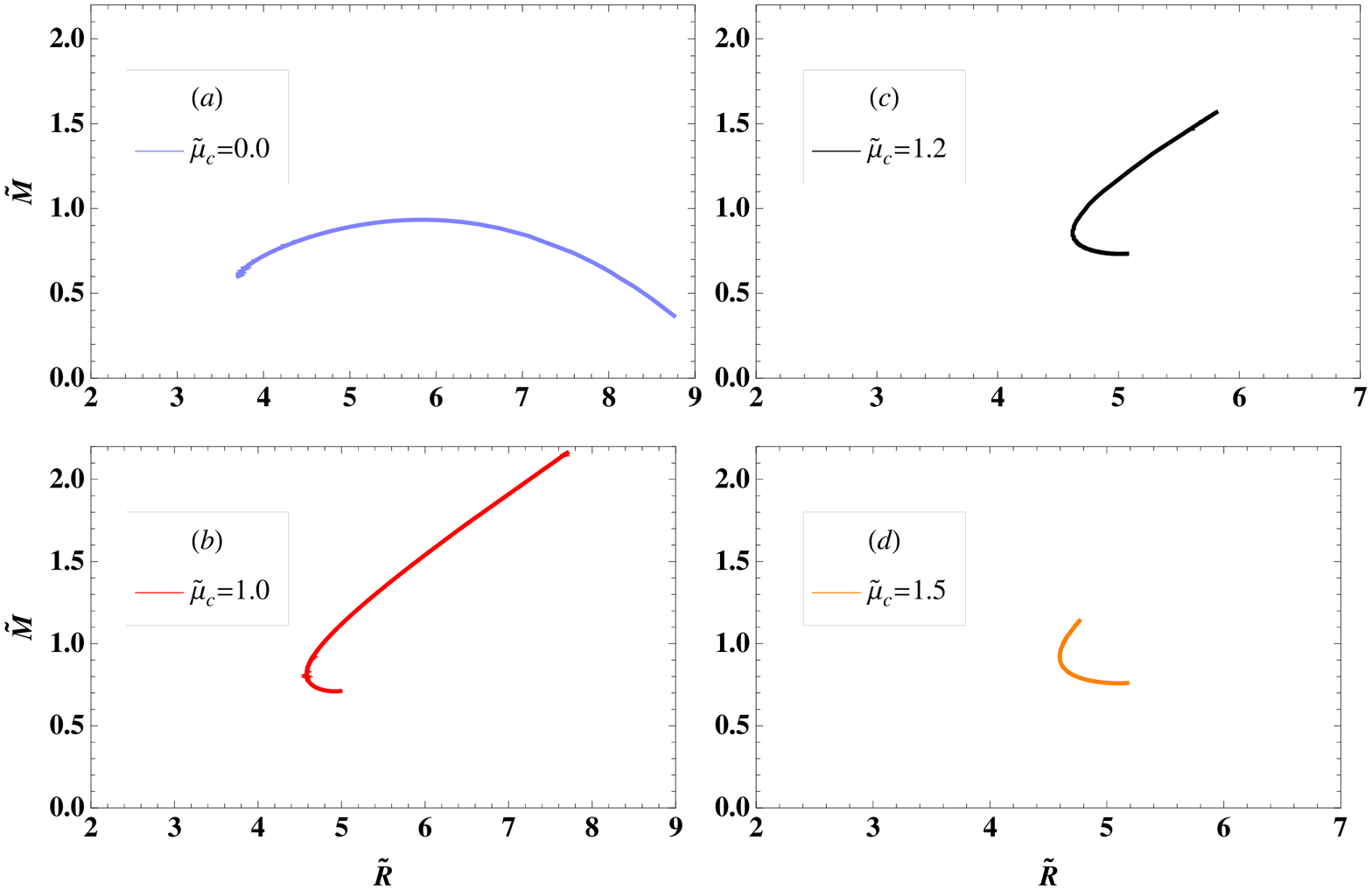}
\caption{Maximum Mass of the star versus Radius for different values of $\tilde{\mu}_c$ is plotted. 
The peak of the graph in each case denotes the highest maximum mass configuration.}\label{MR}
\end{figure}

In figure (4) we have plotted mass of the star versus radius for different values of the
chemical-potential. The radius is determined by the value of radial coordinate $r~(=R)$, where the density  
vanishes. The metric-dependence of the chemical-potential gives a new class of solutions.
Figure (4a) shows the case of zero central chemical-potential, in this case there is no variation
in the chemical-potential as implied by equation (\ref{tov3}). In this case,
the maximum mass value occurs at the peak of the curve and its value
 matches well with the known result \cite{shapiro}.  All the points on the left-hand side of the peak are
unstable, as they have higher gravitational potential energy than the points on the right-hand side.
Figure (4b) represents  $\mu_c=1$ case, one can see that the value of maximum-mass has increased
as compared to $\mu_c=0$ case and more strikingly the value of the maximum-mass does not occur
on the peak of the curve.  The spiral like structure in the lower part of the curve represents two 
branches of then solution. In this region for any given $R$ there exist two points with
the same mass. The points on the lower-arm of the spiral have minimum energy, 
while points on the upper-arm may be unstable. Figure (4c) shows the case with $\mu_c=1.2$.
Again the maximum-mass does not occur on the peak of the curve and its value is smaller
than $\mu_c=1$ case. Radius of the star at the maximum-mass point is much smaller than the
case shown in figure (4b) and thus the density of the star is higher than $\mu_c=1$ case. 
Figure (4d) shows that for $\mu_c=1.5$, the maximum-mass is smaller than $\mu_c=1.2$
case and it occurs at a smaller radius. In this case the maximum-mass point lie in the
unstable region. The lower-arm of the spiral represents the stable branch of the solution.
The maximum-mass point in the stable branch  is comparable to the case shown in figure (4a).
This points occurs when $(\frac{d\tilde{M}}{d\tilde{R}})^{-1}=0 $ is satisfied.
It should be noted that the qualitative arguments for maximum-mass of a star, given below
equation (22),  may not be applicable for the cases shown in figure (4b-4d). 
As these arguments do not account for the variation in the chemical-potential and no-node condition
imposed on the $\rho$.

 Figures (5-6) show  pressure and density as functions of $\tilde{r}$ for the case when the
central-pressure is kept fixed at $\tilde{p}_c=35$ and vary $\tilde{\mu}_c$. Figure (5) indicates
that the pressure decreases less slowly with $r$ by increasing the value of chemical-potential. 
Thus the star radius may increase if the chemical-potential is increasing. The density vs. distance
plot also shows the similar behavior. In this case the value of central-density $\tilde{\rho}_c$ is
not fixed, but it decreases as the chemical-potential at center increases. This behavior is expected
from the form of the EOS (\ref{eos}). 
\begin{figure}
\includegraphics[width=7cm,height=7cm]{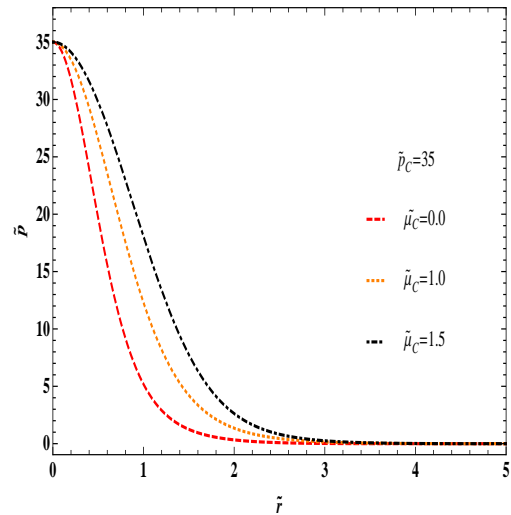}
\caption{ The variation of the normalized pressure with the normalized radius for the
different values of the central chemical-potential is shown here. The central pressure
in all the cases is kept fixed.}
\end{figure}
\begin{figure}
\includegraphics[width=7cm,height=7cm]{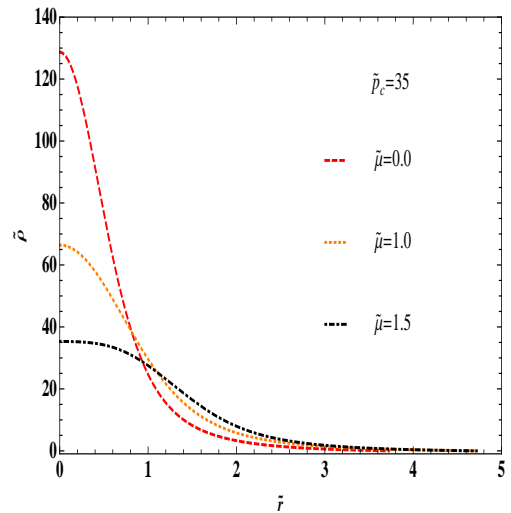}
\caption{ The variation of the normalized density with normalized radius for the different
values of the central-chemical potentials is shown here. The central pressure
in all the cases is kept fixed. }
\end{figure}

Figure (7) shows the typical variation in  chemical-potential 
and the metric-function $B$ with the radius. The metric-function approaches the unity
at the boundary, while the value of the chemical-potential decreases from its central value with
the radius. However, the value of $\tilde{\mu}$ remains non-zero at the boundary.
The dynamics of the chemical-potential and $B$ are complementary as can be anticipated from 
definition of $\tilde{\mu}$.
\begin{figure}
\includegraphics[width=7cm,height=7cm]{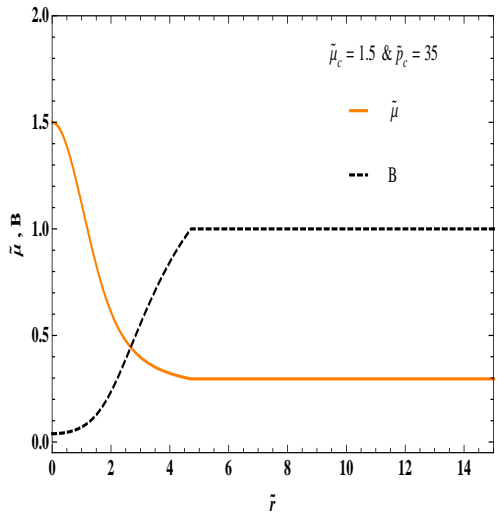}
\caption{The typical variation of the normalized chemical-potential
and the metric-function $B$ with the radius is shown for the given values of
central pressure and chemical-potential.}
\end{figure}

Let us consider the case when the scalar-field represents the diquark-condensates
which may arise in a low-density and strong-coupling regime. The condition for the condensate
formation is $(\mu/m)^2/B>1$ or $\tilde{\mu}^2>1$. One can notice from figure (7) that this
condition is violated somewhere inside the star. For the condition to be valid inside
the entire region of the star one has to increase the value of $\tilde{\mu}_c$.
From the discussion above, increasing $\tilde{\mu}_c$ would also requires to increase the
central-pressure $\tilde{p}_c$. This may in turn increases the density as shown in figure (2).
Clearly the energy density can not increase to a very high values otherwise the 
condition of BEC formation may not be fulfilled. For example one may take the
value of the chemical potential $\tilde{\mu}_c=1.5$ which is consistent with \cite{pir1}.
According to the figure (2), the physical density profile should be around 30 times larger
then the nuclear density which is not consistent with the BCS-BEC crossover assumption.
Moreover, the mass-radius diagrams shown in figure (4) suggest that as $\tilde{\mu}$ increases
the available parameter-space for the stable configuration reduces significantly.
We would like to add here that when one considers the case of fundamental scalar-field, the constraint
of the low energy-density is not required. In this situation the boson-star equilibria are defined
over a large range of the parameters.

\section{SUMMARY}

Thus we have studied the properties of boson-stars in the mean-field approximation.
We derived the generalized set of TOV equations to incorporate the effect of finite
chemical potential. It is shown that chemical potential can introduce unphysical
features like nodes in the equilibrium profile of the density which can be removed
by imposing the no-node condition. No-node condition can put constraints on the physical
values of the chemical potential and the pressure at the center of the star.
The new class of solutions of the star that we have found show that available
parameter space for the stable solutions gets greatly reduced by increasing the
chemical potential. Further when our  analysis when applied to the diquark condensate
stars show that for a stable solution the density of the star needs to be too high for
the diquark condensates to exist.

\acknowledgements
We would like to thank Dr. H. Mishra for the useful discussion.

\end{document}